УДК 001:168.5

*С. Б. Куликов*

## ПРОБЛЕМА ИСТИНЫ В КОНТЕКСТЕ ИНТЕРПРЕТАЦИЙ КВАНТОВОЙ ЛОГИКИ

Доказывается, что анализ проблемы истины в контексте интерпретаций квантовой логики позволяет выйти на более общую проблему связи областей квантово-механических и квантовологических исследований на основе принципа тождественности неразличимых явлений, введенного Г. Лейбницем и развитого С. Крипке. Вместе с тем раскрывается перспектива выявления специфики отношений между квантовой механикой и квантовой логикой в контексте модальных расширений квантовой логики.

**Ключевые слова:** *принцип тождественности неразличимых явлений, квантовая логика, модель, интерпретация.*

Целью статьи является представление результатов анализа проблемы истины в контексте интерпретаций квантовой логики. В более общем ключе вопросы развития квантовой логики затронуты в работе С. Б. Куликова [1]. Поэтому представляемую статью следует расценивать как развитие и уточнение отдельных положений, обсуждавшихся ранее.

Не менее важно отметить еще и такое обстоятельство. Как полагает И. Т. Касавин, в контексте логических исследований раскрывается «отнесенность понятия истины только к логически правильно построенным предложениям естественных и искусственных языков» [2, с. 5]. Из этого ясно, что изучение проблемы истины в контексте интерпретаций квантовой логики предполагает не раскрытие всего спектра гносеологических аспектов проблемы истины, а лишь обнаружение перспектив исследования данного вопроса в его узком смысле.

Под квантовой логикой обычно понимают особую область знаний, начало развития которой положено в трудах Дж. фон Неймана. Идеи этого автора впервые были воплощены Г. Биркгофом. Исходная же суть данных идей заключалась в возможности логических исчислений свойств квантовой системы. В рамках такого понимания квантовая логика суть построение особых формальных образований, в рамках которых выполняются исчисления результатов наблюдения выделенных состояний квантовых систем [3, с. 189; 4, с. 823–843; 5, с. 24, 6, с. 33].

Общая актуальность исследования проблемы истины в контексте интерпретаций квантовой логики обусловлена одним важным обстоятельством. В современной науке все более усиливаются тенденции к установлению междисциплинарных связей. Примером здесь может служить биофизика. В первой половине XX в. Э. Шрёдингер поставил ряд вопросов относительно особенностей реализации физических закономерностей в границах живого вещества [7, с. 11]. Поиск ответов на эти вопросы привел к возникновению направления исследований, в котором начали привлекаться наработки физики, химии, биологии и медицины.

Междисциплинарные связи в естествознании обусловлены объективным единством материального мира. В ходе раскрытия частных сторон материального мира достигается фундаментальный уровень понимания этих сторон. На данном уровне становятся необходимыми пересечения отдельных областей знания. Такие пересечения раскрывают общие зависимости функционирования материальных явлений и процессов от более глубоких закономерностей. Например, биология и этика, образуя комплекс биоэтических знаний, имеют в основе общий интерес к смыслу человеческого присутствия в границах наложения природного мира и сферы общественных отношений [8].

Несколько иначе обстоит дело в области методов исследования квантово-механических процессов, требующих пересечения философии, логики и физики. В философии выделяются принципы, на которых базируется процесс научно-исследовательской деятельности (например, это принцип историзма в исторической науке, принцип природосообразности в педагогическом знании и другие). В логике разрабатываются методы анализа форм мышления, раскрываются законы мыследеятельности в целом. В физике формулируются теоретические положения о природных явлениях и процессах. Эти положения имеют в основном вид математических уравнений, которые обобщают собранные факты. Тем самым фактические данные получают объяснение, а отдельные решения уравнений способны указывать на возможные новые факты. Примерами здесь могут служить два научных открытия: во-первых, фиксация на базе общей теории относительности А. Эйнштейна факта отклонения луча солнца в гравитационном поле, а во-вторых (правда, уже не в физике, а в астрономии), обнаружение Нептуна И. Галле (и его ассистентами) на базе расчетов У. Леверье.

В связи со всем этим исследование логических методов познания квантово-механических процессов, в принципе, должно иметь общие основания. Такие основания необходимы для координации результатов исследовательской деятельности в сфере соответствующих разделов физики, логики и философии. Однако раскрытие общих оснований существенным образом затруднено. Так, В. Л. Васюков замечает, что «в начале 80-х гг. был получен ряд критических результатов относительно некоторых выдвинутых ранее систем квантовой логики, фиксирующих их бесполезность с точки зрения физики» [9, с. 34].

Итак, специфика изучения проблемы истины в контексте интерпретаций квантовой логики, по сути, соответствует постановке задачи по раскрытию оснований для обнаружения связи квантовой логики с действительными процессами квантово-механических исследований.

Авторы полагают, что в ходе решения поставленной задачи особую роль способен сыграть принцип тождественности неразличимых явлений, введенный еще Г. Лейбницем и развитый в особом направлении С. Крипке. Так, С. Крипке считает, что принцип тождественности неразличимых явлений самоочевиден так же, как и принцип недопущения противоречий [10, с. 3]. Это подтверждают исследования в области модальной логики, в частности семантики «возможных слов» как слов, обозначающих возможные ситуации употребления (чрезвычайно любопытные результаты исследований, касающихся данных вопросов, представлены в некоторых работах В. А. Суровцева [11]). В рамках разработок С. Крипке в результате прояснения связей между именованием и необходимостью раскрывается, что контексты вхождения не отображают подлинные качества вещей. В рамках процедур именования в основном выполняется пересечение, а не дополнение смыслов. В итоге требуется строгое разграничение контекстов и придание некоторым понятиям статуса ригидных десигнаторов (под «ригидными десигнаторами» понимаются точные указатели, не меняющие смысла во всех возможных контекстах).

В работе С. Б. Куликова [1] выявлено, что могут быть обнаружены два основных вида точных указателей: партикулярные и универсальные. Причем установлено, что в трудах С. Крипке функции универсально твердого указателя выполняет неразличимость по всем параметрам, выступающая в виде возможности связать конкретные слова общностью смысла. В отдельных же контекстах такие слова могут продолжать употребляться как различные наименования, по сути, одного и того же явления или процесса.

Именно раскрытие самоочевидности принципа тождественности неразличимых явлений, сближение его по статусу с принципом недопущения противоречий позволяет обнаружить способы эффективного решения задачи по раскрытию оснований, на базе которых проясняются связи квантовой логики с действительными процессами квантово-механических исследований.

В этом отношении вслед за результатами, полученными ранее [1], принципиально важно отметить два момента. Связь квантовой логики и принципа тождественности неразличимых явлений обнаруживается, во-первых, в ходе истолкований квантовой логики как расширения модальной логики [12]. Во-вторых, во многом то же обнаруживается при модальных интерпретациях квантовой механики [13]. Из чего видно, что может быть выделена как интердисциплинарная специфика принципа тождественности неразличимых явлений, так и его трансдисциплинарный характер.

Выявление интердисциплинарной (внутрилогической) специфики принципа тождественности неразличимых явлений позволяет обратить внимание на особенность методов, которые применяются в ходе интерпретации квантовой логики как расширения модальной логики. В частности, В. Л. Васюков раскрывает особую последовательность операций. В рамках этой последовательности, во-первых, бесконечнозначная логика Я. Лукасевича интерпретируется как разновидность вероятностной логики. Во-вторых, уже в рамках вероятностной модели Дж. Макки выделяет аксиомы квантовой логики, лежащие в основе квантово-механических экспериментов. В-третьих, Г. Дишкант, в свою очередь, предлагает включить в систему Я. Лукасевича исчисления Дж. Макки [9, с. 57].

Указанные операции выполняются на базе минимум двух методологических приемов: моделирования и интерпретации. Под моделированием понимается воспроизведение свойств объектов и процессов, выраженных в рамках специфической системы абстрактных обозначений. Система обозначений, взятая сама по себе, базируется на совокупности аксиом и позволяет формулировать высказывания, выводимые из аксиом на основе правил вывода. Интерпретация совпадает с такими формулировками и доказательствами некоторых положений, которые позволяют соотнести отдельно взятую предметную область и значение сформулированных положений. В данном отношении нетривиальных результатов достигают А. А. Степанов в процессе интерпретации отдельных аспектов научно-технического творчества, а также И. В. Мелик-Гайказян и И. П. Элентух в границах моделирования способов эффективного решения научно-технических проблем [14, 15].

В то же время в ходе моделирования и интерпретации квантовой механики, например, Дж. Макки замечает: «Нам будет удобно ввести основные понятия квантовой механики аксиоматически. Мы построим строго определенную математическую модель и опишем ее физический смысл настолько точно, насколько это возможно» [16, с. 60]. Из этого ясно, что моделирование и интерпретация предполагаются в виде общих принципов относительно частных методов, применяемых в квантовой логике, а именно аксиоматизации, формализации, анализа, сравнения и других.

В данном отношении, например, согласно Г. Дишканту, введение символа $Q$ и особых модальных правил логического вывода позволяет проинтерпретировать аксиоматику Дж. Макки. Эвристический потенциал этих результатов связан с достижением возможности на базе совместности наблюдений найти подтверждение материальных импликаций. Построенные исчисления раскрывают свою истинность в рамках подведения наблюдений за поведением физических объектов под общие правила квантовой механики.

В. Л. Васюков полагает, что Г. Дишканту в рамках модальной интерпретации правил квантово-механических наблюдений не удалось показать абсолютную семантическую полноту исчислений. Была установлена только относительная полнота этих исчислений сравнительно квантовой пропозициональной логики [9, с. 59]. Более существенных результатов смог достичь Р. Голдблатт. Этот исследователь предложил вариант перевода минимальной квантовой логики с сокращенным числом связок (т. е. ортологики) в одну из версий модальной логики, а именно логики Л. Брауэра [17]. Прояснение связи формул ортологики с многообразием возможных миров открывает перспективу доказательства или опровержения истинности таких миров.

Чрезвычайно важными для данного исследования являются модальные интерпретации квантовой механики, разработанные Б. ван Фраасеном, Д. Диксом и другими исследователями. Эти интерпретации могут быть обобщены в границах различения измеренных и динамических состояний [18]. Открывается перспектива исчисления вероятностных оценок отдельно взятых состояний. Однако в целом квантовая логика не имеет законченной интерпретации. Присутствуют концептуальные ограничения, налагаемые квантовыми формализмами на классические способы понимания, хотя данные ограничения и остаются не вполне очевидными [18, с. 113].

Опора на принцип тождественности неразличимых явлений вносит ясность в вопросы моделирования и интерпретации квантовой логики. Обнаруживаются основания для раскрытия аналогии между квантовыми исчислениями и состояниями квантовых систем. Также оказывается оправданным установление истинности или ложности формальных построений отдельных исчислений.

Итак, анализ проблемы истины в узком ключе в контексте интерпретаций квантовой логики приводит к выводу о том, что логические модели и интерпретации состояний квантовых систем могут строиться на базе особой идеи, а именно идеи связи квантовой механики и принципа тождественности неразличимых явлений. Открывается перспектива выявления специфики отношений между квантовой механикой и квантовой логикой в контексте модальных расширений квантовой логики. Вместе с тем с точки зрения семантики «возможных миров» обнаруживается путь для решения проблемы выявления особенностей построения квантово-механических исследований. Эти исследования получают статус «возможных квантовых механик». Именно потому они могут быть представлены не просто как различение «истины» и «лжи» в рамках суждений о физических явлениях, но и в качестве отображения логически возможных и (или) невозможных, необходимых и (или) случайных и т. д. построений квантово-механических знаний в целом.

## Список литературы

Куликов С. Б., доктор философских наук, доцент, декан, зав. кафедрой.
**Томский государственный педагогический университет.**
Ул. Киевская, 60, Томск, Россия, 634061.
E-mail: kulikovsb@tspu.edu.ru

*S. B. Kulikov*

**TRUTH PROBLEM IN THE CONTEXT OF INTERPRETATIONS OF QUANTUM LOGIC**

The paper defends the thesis that analysis of truth problem in the context of interpretations of quantum logic allows to come to more general problem of the coherence between the scope of quantum mechanical and quantum logical studies on the basis of principle of the indiscernibility of identicals entered by G. Leibnitz and developed by S. Kripke. At the same time it's revealed the prospect of elicitation of specifics of the relations between quantum mechanics and quantum logic in a context of modal expansions of quantum logic.

**Key words:** *principle of the indiscernibility of identical, quantum logic, model, interpretation.*

**Tomsk State Pedagogical University.**
Ul. Kievskaya, 60, Tomsk, Russia, 634061.
E-mail: kulikovsb@tspu.edu.ru